\documentclass{aa}
\usepackage{graphicx}
\usepackage{txfonts}
\usepackage{natbib}
\begin{document}
   \title{Sunspot group tilt angles and the strength of the solar cycle}

   %\subtitle{}

   \author{M. Dasi-Espuig\inst{1},
	  S. K. Solanki\inst{1,2},
          N. A. Krivova\inst{1},
          R. H. Cameron\inst{1},
          \and
          T. Pe\~{n}uela\inst{1,3,4}
%          C. Ptolemy\inst{2}\fnmsep\thanks{Just to show the usage
%          of the elements in the author field}
          }

   \offprints{M. Dasi-Espuig}

   \institute{Max-Planck-Institut f\"ur Sonnensystemforschung,
              Max-Planck-Str. 2, 37191 Katlenburg-Lindau, Germany\\
              \email{dasi@mps.mpg.de}
          \and
              {School of Space Research, Kyung Hee University, Yongin, Gyeonggi, 446-701, Korea\\
             }
         \and
             Astronomical institute of the St. Petersburg State University,
             Universitetsky prospekt, 28, 198504, Peterhof, St. Petersburg, Russia.\\
         \and
             European Southern Observatory, ESO, 
             Karl-Schwarzschild-Strasse 2 D-85748,
             Garching bei München
            }

   \date{Received ; accepted }

% \abstract{}{}{}{}{} 
% 5 {} token are mandatory
 
  \abstract
   {It is well known that the tilt angles of active regions increase with their latitude (Joy's law). It has never been checked before, however, whether the average tilt angles change from one cycle to another. Flux transport models show the importance of tilt angles for the reversal and build up of magnetic flux at the poles which is, in turn, correlated with the strength of the next cycle.}
    {Here we analyse time series of tilt angle measurements and look for a possible relationship of the tilt angles with other solar cycle parameters, in order to glean information on the solar dynamo and to estimate their potential for predictions of solar activity.}
   {We employ tilt angle data from Mount Wilson and Kodaikanal observatories covering solar cycles 15 to 21. We analyse the latitudinal distribution of the tilt angles (Joy's law), their variation from cycle to cycle and their relationship to other solar cycle parameters, such as the strength (or total area covered by sunspots in a cycle), amplitude and length.}
   {The two main results of our analysis are: 1. We find an anti-correlation between the mean normalized tilt angle of a given cycle and the strength (or amplitude) of that cycle, with a correlation coefficient of $r_{c}=-0.95$ (99.9\% confidence level) and $r_{c}=-0.93$ (99.76\% confidence level) for Mount Wilson and Kodaikanal data, respectively. 2. The product of the cycle averaged tilt angle and the strength of the same cycle displays a significant correlation with the strength of the next cycle ($r_{c}=0.65$ at 89\% confidence level and $r_{c}=0.70$ at 92\% confidence level for Mount Wilson and Kodaikanal data, respectively). An even better correlation is obtained between the source term of the poloidal flux in Babcock-Leighton-type dynamos (which contains the tilt angle) and the amplitude of the next cycle. Further results are: We confirm the linear relationship (Joy's law) between the tilt angle and latitude with slopes of 0.26 and 0.28 for Mount Wilson and Kodaikanal data, respectively. In addition, we obtain good positive correlations between the normalized area weighted tilt angle and the length of the following cycle, whereas the strength or the amplitude of the next cycle do not appear to be correlated to the tilt angles of the current cycle alone.}
   {The results of this study indicate that in combination with the cycle strength, the active region tilt angles play an important role in building up the polar fields at cycle minimum.}

   \keywords{Sunspots -- Sun: dynamo -- Magnetic fields -- Sun: activity
               }
   \authorrunning{Dasi-Espuig et al.}
   \titlerunning{Sunspot tilt angles and strength of the solar cycle}

   \maketitle
%
%________________________________________________________________

\section{Introduction}

Solar cycles differ from each other, showing different lengths, amplitudes and strengths. Understanding the cause of such variations and, ideally, reproducing them is one of the aims of dynamo theory.

Magnetic flux transport dynamo models of the Sun's global magnetic field have been shown to reproduce fairly well the amplitude and duration, among other characteristics, of the solar cycle for at least a few cycles \citep{charbonneau05, charbonneau07, dikpati06}. Some of the key ingredients of such models include differential rotation, meridional flow, latitude distribution of sunspots, latitudinal drift and a systematic tilt angle of the bipolar groups (Joy's law). These ingredients together explain the polarity reversal of the magnetic field at the poles every $\sim11$ years. Due to differential rotation the magnetic field lines are wound up around the Sun's rotation axis and when this field is strong enough it becomes buoyant and rises to the surface as sunspots \citep{babcock61,dikpati08}. The magnetic flux from the sunspots is carried by the meridional flow to the poles, finally causing the reversal. It was already proposed by \citet{leighton69} that for the reversal to occur there must be cancellation between the leading portions of spots on opposite hemispheres through the slight tilt of the bipolar regions. In this way a greater fraction of the following polarity flux reaches the poles.

\citet{schrijver02} tested the hypothesis where the polar magnetic field on the Sun is determined by the accumulation of field transported poleward from sunspots at lower latitudes as a consequence of the tilt in the bipoles. Their model was not able to reproduce the polar field measurements of the past years if only the rate at which sunspots emerge is varied from one cycle to another. Furthermore, \citet{wang02} included a cycle to cycle variable meridional flow in order to achieve agreement with other observations, demostrating that this variable meridional flow could serve as a regulator of the polarity reversal process. Flux-transport simulations also showed that the strength of the polar fields, which feed the dynamo and help determine the strength of the next cycle, is sensitive to the average tilt angle of the active regions of the previous cycle \citep{baumann04}. Here we investigate whether there is a variation of the cycle averaged tilt angle and Joy's law from cycle to cycle and whether there is a relationship between the tilt angles and the strength, i.e. the activity level, of the following cycle.

Previous studies of the sunspot tilt angles have mainly focused on their relationship with other spot parameters such as magnetic flux, drift motions, rotation, area, polarity separation and cycle phase among others \citep{howard91-a, howard96-a, sivaraman07}. Variations from one activity cycle to the next, however, have never been explored. This could shed some light on the mechanism by which the magnetic field of active regions is transported to the poles and may thus have the potential for forecasting future solar activity.\\

Prediction of future solar activity is one of the main challenges in solar physics and is not only of scientific importance but potentially helps to make predictions about changes to our natural environment that can affect our lives, e.g. space weather and Earth's climate. Most of the present day predictions are based on statistical analyses of solar activity in the past \citep{hathaway99,hathaway09}. A more physics-based approach is offered by models of the evolution of the Sun's magnetic field, although recent dynamo computations have given controversial results for cycle 24 \citep{dikpati06, choudhuri07,jiang07}.

The paper is structured as follows. In Sect. 2 we describe the data, the method and some tests. Section 3 presents the results, which are then discussed in Sect. 4. In Sect. 5 we present our main conclusions.

%__________________________________________________________________

\section{ Data and Tests }

  \subsection{ Data }

For our analysis we employ sunspot data derived from white light images taken at Mount Wilson and Kodaikanal observatories. These observatories have regularly observed the solar disc in white light since the beginning of the 20th century. The data we use cover the years 1917 to 1985 and 1906 to 1987 for Mount Wilson and Kodaikanal, respectively. This means that cycles 15 to 21 are completely covered by the Kodaikanal (hereafter KK) record, but the first 4 years of cycle 15 and the last year of cycle 21 are missing in the Mount Wilson (hereafter MW) series. \citet{howard84} measured the positions and areas of individual sunspots on digitized MW images and then grouped the sunspots using a technique based on proximity. The grouping of individual sunspots was done by \citet{howard84} by applying a running box, 3$^\circ$ wide in latitude and 5$^\circ$ wide in longitude, centered at each spot on the solar disc. Any other spot that fell inside the box was included as part of the group. To distinguish between the leading and following portions of the sunspot groups, they first computed the mass center. The portion to the east of the mass center was defined as the leading portion and the portion to the west as the following. This definition was applied to all sunspot groups since they had no magnetic information \citep{howard91-a}. The tilt angle of a sunspot group, $\alpha$, is defined as $\tan\alpha=\Delta\phi/ [\Delta l\cos\phi]$, where $\phi$ is the latitude of the center of the sunspot group and $\Delta\phi$ and $\Delta l$ are the differences in latitude and central meridian position between the center of gravity of the leading and following portions of the group, respectively.

The final data set includes dates of observations, positions, area and number of individual sunspots for each sunspot group and for its leading and following portions, as well as the tilt angle. A description of the calculation of the tilt angles can be found in the paper by \citet{howard91-a}. The white light images from Kodaikanal observatory were treated using the same techniques and procedures by \citet{sivaraman93}.

In the present study we also use the sunspot area data set compiled by \citet{balmaceda09} using observations from a number of different observatories that were carefully cross-calibrated in order to reduce, as much as possible, systematic and other differences between observations at different sites as well as the number of data gaps. The combined data set goes back to 1874.

%figura de polarity separation
   \begin{figure}[h]
   \flushleft
   \includegraphics[scale=0.5]{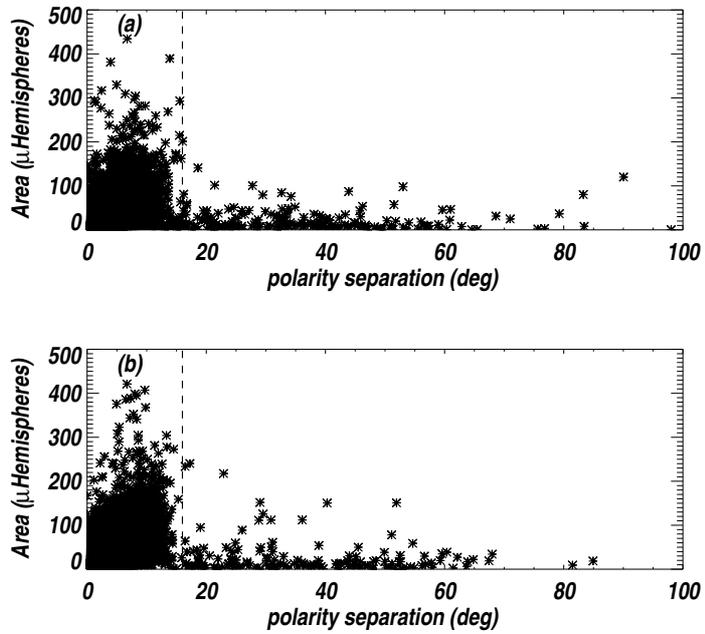}
      \caption{Area versus polarity separation between the leading and following portions of a sunspot group for (a) Mount Wilson and (b) Kodaikanal data sets. The dashed line corresponds to $16^\circ $. }
      %   \label{FigVibStab}
   \end{figure}

  \subsection{ Data evaluation }

The first inspection of the MW and KK data revealed a great number of zero values for the tilt angles ($\sim$ 22\% in MW and $\sim$ 30\% in KK). Therefore, we checked whether all zeros are real or just missing values. We assumed that the tilt angle can only be determined if the leading and following portions of a group have at least one spot each. Thus we neglected the zero tilt angle values in all cases when either portion of a group contained no spots. By applying this criterion, we found only one real measurement of a tilt of exactly zero degrees in the MW data set and none in the KK data set. The rest of the zero values only mark that it was not possible to measure the tilt angles for some reason. In order to accept a tabulated tilt angle as valid, we also required that the distance between the leading and following portions is less than $16^\circ $. This is justified by the distribution of sunspot group areas with polarity separation (see Fig. 1). Most of the groups lie in a range between 0 -- 350 microhemispheres in area and $0^\circ-16^\circ$ in polarity separation. Only around 0.6\% and 0.4\% of all the groups in the MW and KK data sets, respectively, present a polarity separation bigger than $16^\circ$. Of these, most have areas below 70 microhemispheres and are thus most likely typos since their total area is small and at such polarity separations it is most improbable that the two polarities belong to the same group. All together, i.e. based on both criteria, we have rejected $22.5\%$ of MW groups and $30.6\%$ of KK ones.

%figura de dlat, dlong
   \begin{figure}[h]
   \flushleft
   \includegraphics[width=0.5\textwidth]{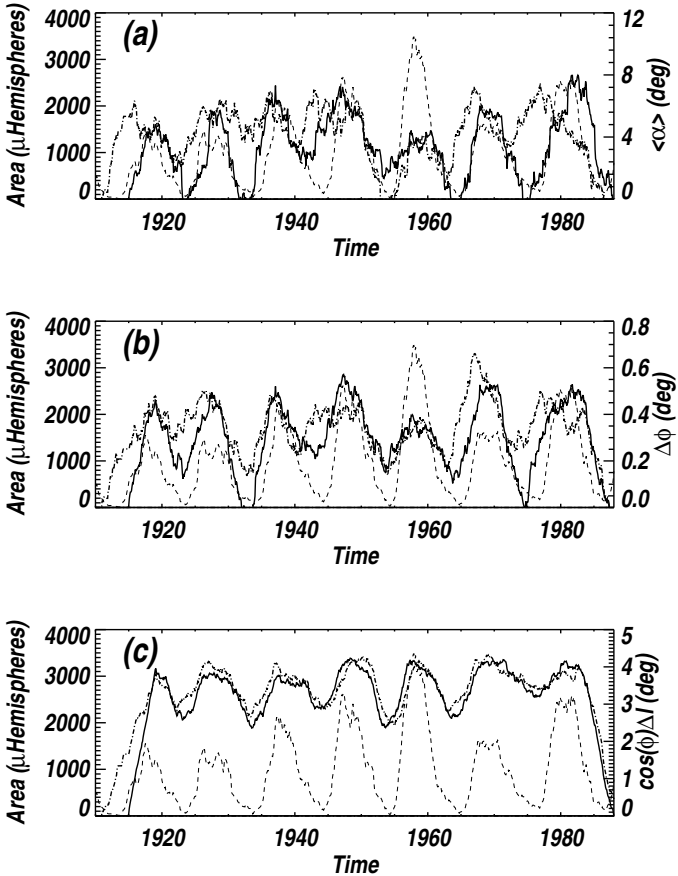}
      \caption{In all panels: Dashed thin line and left-hand Y-axes show the monthly means of the sunspot area from \citet{balmaceda09} vs. time; the solid and the dot-dashed thick lines represent Mount Wilson and Kodaikanal data, respectively. (a) Monthly area-weighted means of sunspot tilt angle smoothed over 4 years (right-hand Y-axis), (b) area-weighted latitude separation (right-hand Y-axis) and (c) area-weighted longitude separation of leading and following parts of sunspot groups (right-hand Y-axis).
              }
       %  \label{FigVibStab}
   \end{figure}

In our analysis we considered mainly cycle averages to study the variaton of the tilt angles from cycle to cycle. The values of these averages are not greatly affected by gaps in the measurements if these are distributed homogeneously throughout each cycle. If, however, gaps are, for example, dominantly found in the ascending phase of a solar cycle the mean value of the tilt angle for that cycle will be lower than the real value. This is a direct consequence of the butterfly diagram (spots at the beginning of a cycle appear at higer latitudes) and Joy's law (tilt angles are higher for sunspot groups located at higher latitudes). Unfortunately it is not possible to distinguish between spotless days and gaps in the data sets of MW and KK observatories. We find that no spots were present or no observations were made on $\sim60\%$ and $\sim55\%$ of all days in the MW and KK data sets respectively. In fact most of them are located within solar activity minima, which is reasonable because the number of truly spotless days is largest at minimum activity. It was possible to determine the spotless days by comparing the MW and KK data sets with a more complete daily sunspot data set from \citet{balmaceda09}. In this data set, where it is possible to distinguish between gaps and spotless days, only $6.5\%$ of the data are missing due to gaps in the considered period. After the comparison we retrieved a $\sim59\%$ and a $\sim56\%$ of truly missing dates in MW and KK data sets respectively. These remaining gaps seem on average to be more or less randomly distributed over cycle phases. They should not significantly affect the averages over a cycle. However the gaps do affect the calculation of the cycle length and strength. In our work we use the length as defined and calculated by the National Geophysical Data Center (see Section 3.2). In the case of the strength, we used the daily sunspot area data set from \citet{balmaceda09} due to its low percent of gaps. To avoid systematic errors, we linearly interpolated across the sunspot area data gaps.

Next we consider another possible source of bias. Since during a strong cycle more spots are found on the Sun's surface the grouping criterion by Howard ($3^\circ$ wide in latitude and $5^\circ$ wide in longitude) could lead to an erroneous grouping for such cycles, because spots that do not belong to the group, but rather to a nearby neighbouring region, could be included. We expect such misclassification to occur mainly in the longitudinal direction due to the asymmetry of the box. It would lead to an enhancement of the longitudinal separation between the following and leading portions of the sunspot groups ($\Delta l\cos\phi$) for the strongest cycles. Such a spuriously increased longitude separation would lead to a reduced average tilt. Figure 2(a) shows the time series of monthly means of sunspot group tilt angles weighted by their corresponding sunspot group areas (see right-hand Y-axis) for both MW (solid line) and KK (dot-dashed line) records. A smoothing of 4 years was necessary due to the high noise the data presented.

The mean area-weighted tilt angles are calculated as follows:
\begin{displaymath}
  \langle\alpha_{\omega}\rangle=\frac{\Sigma A_{j}\alpha_{j}}{\Sigma A_{j}},
\end{displaymath}
where $A_{j}$ and $\alpha_{j}$ are the area and the tilt angle of the sunspot group $j$, respectively. In the case of monthly means, the sum goes over all sunspot groups in one month, while in the case of cycle means, the sum goes over all sunspot groups present during one cycle. Also plotted are monthly means of sunspot area (dashed line and left-hand Y-axes) from \citet{balmaceda09}. Cycle 19 is clearly the strongest and has, at the same time, the lowest values of the tilt angles of the 7 analysed cycles. We test in Fig. 2 (b) and (c) the possibility of a systematic error. Here we plot monthly means of sunspot latitude ($\Delta\phi$) and longitude ($\Delta l\cos\phi$) separations between the leading and following portions of sunspot groups. The solid line is again used for the data from MW observatory and the dot-dashed line for KK observatory. The drop in $\Delta\phi $ during cycle 19 indicates that the low values of the tilt angles during this period are due to a lower latitudinal separation of both polarities while the fact that cycle 19 is not conspicuous in Fig. 2 (c) indicates that its low tilt angles are not due to a larger longitudinal separation. This result does not exclude that the grouping algorithm might have combined together sunspots that with magnetic information would have been grouped differently. In any case we believe that, if there is a systematic error in the tilt angles, this is not seen in the latitude and longitude positions of the leading and following portions of sunspot groups.

\subsection{Joy's Law}

The tilt angle dependence on the latitude was first found by Joy in 1919 \citep{hale19} and later confirmed by other authors \citep{howard91-a, wang91, sivaraman99}. It provides strong constraints on the magnetic field strength of the flux tubes in the tachocline, which emerge to form the observed active regions \citep{dsilva93, schussler94}. This relation shows a positive trend: the tilt angles are larger for sunspots at higher latitudes. We have used data from MW and KK observatories in order to rederive this relationship as a test. In Fig. 3, the tilt angles averaged over the complete data sets for latitude bins of $5^\circ$ are plotted versus latitude. MW data are represented by asterisks connected by the dashed line and KK observations by open circles connected by the solid line. The results for both data sets are in agreement with each other within $1-2\sigma$ and are very close to the results obtained by \citet{sivaraman99}, with the mean difference between our points and theirs being $|\Delta|\sim0.1^{\circ}$ and $|\Delta|\sim0.2^{\circ}$ for MW and KK, respectively. These differences are most probably due to different selection criteria applied to the data.

%figura de joy's law
   \begin{figure}[h]
   \flushleft
   \includegraphics[width=0.5\textwidth]{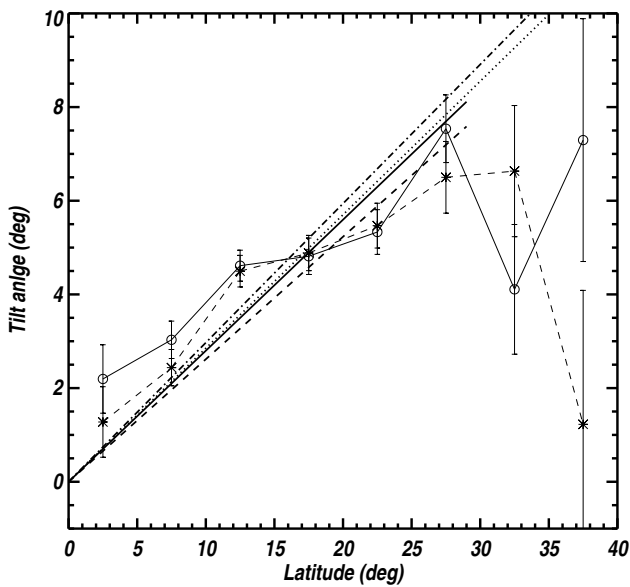}
      \caption{Mean tilt angle for bins of 5 degrees latitude
               vs. latitude for MW (asterisks connected by the dashed line) and
               KK (open circles connected by the solid line). The error bars represent
		$\pm 1$ standard error. The thick dashed and
               solid lines are linear fits forced to pass through the origin to the MW and KK data, respectively. The dotted and dot-dashed lines correspond to lines where the slope has been taken as the ratio of tilt with latitude (see description in the text) for the MW and KK data respectively.
              }
        % \label{FigVibStab}
   \end{figure}

We have also obtained linear fits to the data points excluding the last two bins since the number of groups in these is very low compared to the other bins (about 2\% and 0.5\% of the total) and the errors are higher by a factor of $\geq2$. The fits are forced to pass through the origin since we expect no tilt for sunspot groups at the equator. The values for the fits are:
\begin{displaymath}
  \alpha=(0.26\pm0.05)\lambda
\end{displaymath}
and
\begin{displaymath}
 \alpha=(0.28\pm0.06)\lambda
\end{displaymath}
for MW and KK data, respectively, where $\alpha$ represents the tilt angle (in degrees) and $\lambda$ the latitude (in degrees). The correlation coefficient of the regression lines are $r^{2}=0.85$ for the MW data and $r^{2}=0.76$ for the KK data. The linear fits are shown in Fig. 3, where the thick dashed line is for the MW fit and the thick solid line for the KK fit. Fitting the data points while taking into account that each point has a different value for the standard deviation gives the same result for the slope up to the third decimal. Also plotted are the lines $\alpha=M\lambda$ where $M$ is calculated as $M=\Sigma_{j}\alpha_{j}/\Sigma_{j}\lambda_{j}$ and $\alpha_{j}$ and $\lambda_{j}$ correspond to the tilt angle and latitude of sunspot group $j$. The dotted line represents the MW data with a slope of $M=0.29$ and the dot-dashed line the KK data with a slope of $M=0.30$. This is shown as comparison since it is not affected by the fact that each bin contains a different number of points.

The slopes found here are lower than those obtained by \citet{wang89, wang91} from daily magnetograms for cycle 21 ($\sin\gamma=0.48\cos\theta+0.03$ where $\gamma$ is the tilt angle and $\theta$ the colatitude). However, our values for the slope, 0.26 and 0.28, are closer to the 0.15 value deduced by \citet{schussler06} who used a flux-transport model to fit data from MW and Wilcox Solar obervatories of the total photospheric field. The difference to the results of \citet{wang89, wang91} could have a variety of causes, such as the different types of data considered (spots vs. magnetograms), differences in spatial resolution \citep[cf.][]{howard84,wang89} combined with the dependence of tilt angle on the size of a region \citep{dsilva93-b, howard93}, or the fact that they considered a single cycle. Our results for the slope of the regression line show considerable scatter from cycle to cycle, even to the extent that we do not consider the values obtained for individual cycles to be particularly reliable (see Sect. 3.1).

%______________________________________________________________

\section{Results}

 \subsection{Average value of tilt angles}

It was first pointed out by \citet{howard91-a} that the average tilt angle of all sunspot groups during the period 1917 -- 1985 deduced from MW data was $4.2^\circ\pm0.2^\circ$. For the whole MW data set period (1917 -- 1985) we obtained a value of $4.25^\circ\pm0.18^\circ$, in excellent agreement with \citet{howard91-a}, while for the whole KK data set (1906 -- 1987), we deduced $4.51^\circ\pm0.18^\circ$.

 \begin{table}[h]
 \caption{ Area-weighted mean tilt angles in degrees for each cycle for MW and KK records. }             % title of Table
 \label{table:1}      % is used to refer this table in the text
 \centering                          % used for centering table
 \begin{tabular}{c c c}        % centered columns (4 columns)
 \hline\hline                 % inserts double horizontal lines
 Cycle & MW $\pm$ $1\sigma$ & KK $\pm$ $1\sigma$  \\    % table heading 
 \hline                        % inserts single horizontal line
    15 & 5.69 $\pm$ 0.57 & 5.00 $\pm$ 0.50 \\      % inserting body of the table
    16 & 5.08 $\pm$ 0.46 & 5.91 $\pm$ 0.43 \\
    17 & 5.83 $\pm$ 0.42 & 6.41 $\pm$ 0.41 \\
    18 & 5.69 $\pm$ 0.35 & 4.97 $\pm$ 0.38 \\
    19 & 3.84 $\pm$ 0.33 & 4.59 $\pm$ 0.38 \\
    20 & 4.63 $\pm$ 0.38 & 5.73 $\pm$ 0.36 \\
    21 & 5.30 $\pm$ 0.40 & 5.37 $\pm$ 0.42 \\
 \hline                                   %inserts single line
 \end{tabular}
 \end{table}

Next we treat MW and KK data on a cycle-by-cycle basis, obtaining a different value of the average sunspot tilt angle for each cycle. Figure 2(a) displays monthly area-weighted means of sunspot tilt angles smoothed over 4 years through cycles 15 to 21 for the MW and KK data sets. Table 1 gives area-weighted cycle averages and 1$\sigma$ standard error. Note that cycles 15 and 21 are not complete in the MW data set, as discussed in Sect. 2.1, and thus the value of the mean tilt angle for cycle 15 might be underestimated and the value for cycle 21 overestimated, according to the combination of Maunder's butterfly diagram and Joy's law. The low value of the sunspot tilt angles for cycle 19, as compared to the other cycles, indicated by Fig. 2a, is also seen in the cycle averaged values.

In addition to the average tilt angles, we also attempted to determine Joy's law per cycle in the same manner as done in Sect. 2.3. However, the scatter of the individual values of the mean tilt angle per $5^\circ$ latitude bins turn out to be too large. The correlation coefficient for a linear regression to the points are for some cycles as low as $r^{2}=0.17$ for MW and $r^{2}=0.026$ for KK. Also, the errors in the calculated slopes are comparable to or slightly bigger than the difference between these values from MW to KK data sets. Therefore, no clear difference could be determined between the slopes of Joy's law from cycle to cycle.

 \subsection{Cycle parameter definitions}

For the parameter study we focus on three main characteristics of a solar cycle: strength, amplitude and length. Strength is defined as the total surface area covered by sunspots throughout a given solar cycle. We calculate it from the daily sunspot area data set compiled by \citet{balmaceda09} as the integral of sunspot area over the duration of each cycle. This record is used since it has significantly fewer data gaps than the MW and KK data sets, as discussed in Sect. 2.2. The cycle amplitude is the highest value of monthly averaged sunspot number and the length is the period of time between two consecutive minima. Times of solar activity minimum, amplitudes and the lengths of cycles are taken from the National Geophysical Data Center; http://www.ngdc.noaa.gov/stp/SOLAR/getdata.html.

We looked for possible relationships of these parameters with four different quantities based on the tilt angles: cycle mean tilt angle, $\langle \alpha \rangle$, cycle mean tilt angle normalized by the mean latitude of sunspots during that cycle, $\langle \alpha \rangle/\langle \lambda \rangle$, cycle mean area-weighted tilt angle, $\langle \alpha_{\omega} \rangle$, and the cycle mean area-weighted tilt angle normalized by the mean latitude of sunspots during the same cycle, $\langle \alpha_{\omega} \rangle/\langle \lambda \rangle$. (For a brief discussion of how these choices are influenced by the scatter in the tilt angles see Appendix A).
The area-weighted tilt angles are used to give more importance to the bigger groups, which exhibit less scatter, and the normalized tilt angles are considered in order to remove the effect of the latitudinal dependence (Joy's law) on the cycle-averaged (area-weighted) tilt angles. 
Note that for the MW data set, cycles 15 and 21 are not taken into account in the relationships concerning $\langle \alpha \rangle$ and $\langle \alpha_{w} \rangle$ due to their incompleteness and could be thus biased by Joy's law. This is not the case for the quantities $\langle \alpha \rangle/\langle \lambda \rangle$ and $\langle \alpha_{\omega}\rangle/\langle \lambda \rangle$ since normalizing by the mean latitude removes this source of bias.
Sunspots in stronger cycles lie at higher latitudes \citep{solanki08}, so that simply due to Joy's law these cycles would have larger mean tilt angles. Dividing by the mean latitude largely removes this difference (both, Joy's law and the dependence of mean latitude on cycle strength are linear), so that $\langle \alpha \rangle/\langle \lambda \rangle$ and $\langle \alpha_{\omega}\rangle/\langle \lambda \rangle$ indicate intrinsic changes of Joy's law from cycle to cycle.

 \subsection{Relationships within the same cycle}

We first investigate the possible relationship of the cycle averaged sunspot tilt angles with the three solar cycle parameters of the \textit{same} cycle. These relations may help to shed light on the underlying magnetic flux tubes at the base of the convection zone and the processes that affect them on their way to the surface (in the case of the strength and amplitude of the cycle) and on the possibility that the tilt angles of active regions are involved, along with other features (e.g. meridional flow), in the regulation of the cycle period of the dynamo (in the case of length), or conversely are influenced by it.

%TABLA CON R'S DEL MISMO CICLO--> MW
\begin{table*}
%\begin{minipage}[t]{\columnwidth}
\caption{Correlation coefficients between the 4 quantities based on the tilt angle and the strength ($S$), amplitude ($A$) and length ($L$) of the same cycle.}      % title of Table
\label{table:2}      % is used to refer this table in the text
\centering                          % used for centering table
\renewcommand{\footnoterule}{}  % to avoid a line before footnotes
 \begin{tabular}{c c c c c c c| c c c c c c c}  % centered columns (7 columns)
\hline\hline                 % inserts double horizontal lines
  & \multicolumn{6}{c|}{Mount Wilson} &  \multicolumn{6}{c}{Kodaikanal} \\
\hline\hline
Parameter & \multicolumn{2}{c}{$S$} & \multicolumn{2}{c}{$A$} & \multicolumn{2}{c|}{$L$} & \multicolumn{2}{c}{$S$} & \multicolumn{2}{c}{A} & \multicolumn{2}{c}{L} \\
\hline                        % inserts single horizontal line
       &  $r_{c}$  &  $P$   &  $r_{c}$  &  $P$   &  $r_{c}$  &    $P$  &  $r_{c}$  &  $P$   &  $r_{c}$  &  $P$   &  $r_{c}$  &    $P$  \\
\hline
$\langle \alpha \rangle$                  & $-0.59$ & 0.30 & $-0.60$ & 0.28 & $-0.29$ & 0.64 & $-0.77$ & 0.04        & $-0.69$ & 0.09        & $-0.58$ & 0.17 \\
$\langle \alpha_{\omega} \rangle$         & $-0.48$ & 0.41 & $-0.48$ & 0.41 & $-0.46$ & 0.44 & $-0.46$ & 0.30        & $-0.66$ & 0.11        & $0.19$  & 0.68 \\
\hline
$\langle \alpha \rangle/\langle \lambda \rangle$    & $-0.95$ & $1\times10^{-3}$ & $-0.83$ & 0.02 & $-0.40$ & 0.37 & $-0.93$ & $2\times10^{-3}$ & $-0.82$ & 0.02        & $-0.48$ & 0.30 \\
$\langle \alpha_{\omega} \rangle/\langle \lambda \rangle$ & $-0.81$ & 0.03 & $-0.91$ & $4\times10^{-3}$ & $0.08$ & 0.86 & $-0.80$ & 0.03        & $-0.91$ & $4\times10^{-3}$ & $0.03$  & 0.95 \\
\hline                
 \end{tabular}
%\end{minipage}
\parbox[b]{21cm}{Correlation coefficients are represented by $r_{c}$ and the probability that the correlation is due to chance by $P$ for both the MW and KK data sets.}
\end{table*}

%alpha/lat vs. strength
   \begin{figure}[h]
   \flushleft
   \includegraphics[scale=0.5]{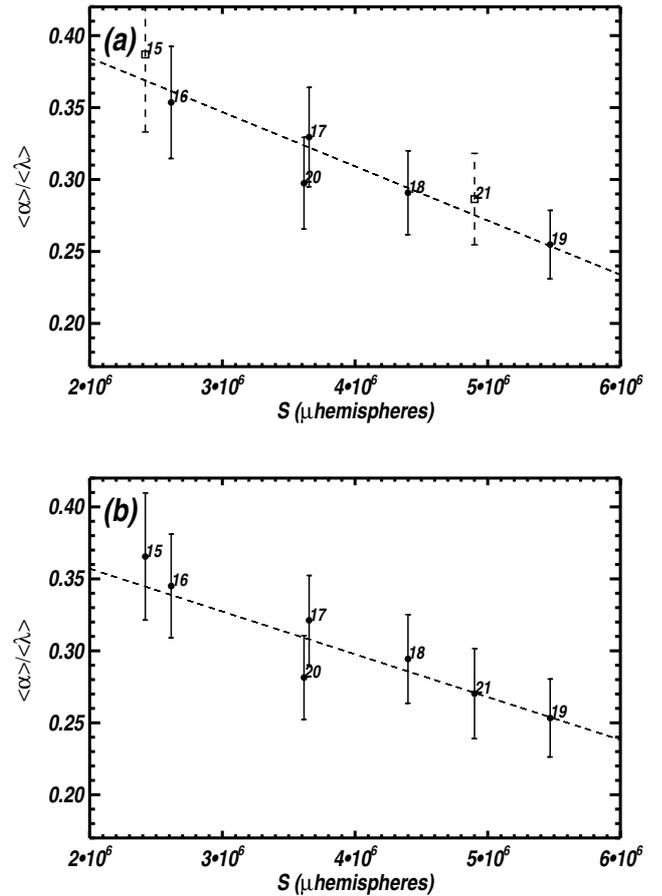}
      \caption{Cycle averaged tilt angle normalized by the emergence latitude vs. strength of the same cycle. The error bars represent $1\sigma$ errors and the dashed line is a linear fit to the points. Panel (a) displays the results based on MW data ($r_{c}=-0.95$) , where cycles 15 and 21 are shown as squares and dashed lines for the error bars, and panel (b) on the KK data set ($r_{c}=-0.93$).}
       %  \label{FigVibStab}
   \end{figure}

We calculated linear correlation coefficients between the 3 solar cycle global parameters and the 4 quantities based on the tilt angles (see Sect. 3.2). Due to the low number of cycles, we also determined the probability that the correlations are due to chance ($P$). These are calculated from the probability density function of the student's t-distribution, which depends both on the correlation coefficient and the number of points in the sample. All the values are listed in Table 2 for MW and KK data. Table 2 suggests that both the strength and the amplitude of a cycle show a significant negative correlation with the average tilt of the same cycle, $\langle \alpha \rangle$, for at least KK data.
For MW data, the probabilities that the correlations are due to chance, $P$, are about $30\%$, but for KK data (that includes both cycles 15 and 21), the corresponding probabilities are lower than $10\%$.
These correlations are significantly strengthened once we eliminate the enhanced effect of Joy's law on cycles with sunspots on average at higher latitudes by considering $\langle \alpha \rangle/\langle \lambda \rangle$.
The probabilities then fall to values below $2\%$ for both MW and KK data sets. For the area-weighted tilt angles, the correlation coefficients are weaker. Although these are also strengthened after the normalization by $\langle \lambda \rangle$, reaching probability values below $3\%$, they remain slightly higher than for $\langle \alpha \rangle/\langle \lambda \rangle$. The correlations between the length and the 4 tilt angle based parameters are in general low, of low confidence and inconsistent in sign between the two data sets.

Figure 4 shows $\langle \alpha_{i} \rangle/\langle \lambda_{i} \rangle $ versus $S_{i}$, where $i$ is the cycle number. The dashed line represents a linear fit to the points and the error bars correspond to $1\sigma$ errors calculated by means of error propagation, where the errors for the mean tilt angle and the mean latitude correspond to their standard error. The error bars have been calculated assuming Gaussian statistics and are thus overestimated. In MW data (Fig. 4(a)) cycles 15 and 21 are represented by squares and dashed lines for the error bars to denote their incompletenes. Note that all data points lie roughly within $1\sigma$ of the regression lines. This suggests that given the accuracy of the measured tilt angles (given largely by the scatter shown by active regions) the obtained correlation coefficients are near the maximum value achievable for data with such large uncertainty.

 \subsection{Relationships with the following cycle}

Prediction of future solar activity is important not only for space weather and climate, but also to test current dynamo models. In this section we investigate how the cycle averaged tilt angles are related to the global parameters of the \textit{next} cycle. We calculated the correlation coefficients between the tilt-angle parameters $\langle\alpha\rangle$, $\langle\alpha\rangle/\langle\lambda\rangle$, $\langle\alpha_{\omega}\rangle$ and $\langle\alpha_{\omega}\rangle/\langle\lambda\rangle$ of cycle $i$ and the parameters $S$, $A$ and $L$ of cycle $i+1$ and the probability that these correlations are due to chance.

%TABLA CON R'S DEL CICLO SIGUIENTE--> (SOLO LENGTH)
\begin{table}[h]
\begin{minipage}[t]{\columnwidth}
\caption{Correlation coefficients between the 4 quantities based on the tilt angle and the length, $L$, of the next cycle.}      % title of Table
\label{table:3}      % is used to refer this table in the text
\centering                          % used for centering table
\renewcommand{\footnoterule}{}  % to avoid a line before footnotes
\begin{tabular}{c c c| c c c}  % centered columns (7 columns)
\hline\hline                 % inserts double horizontal lines
  & \multicolumn{2}{c|}{Mount Wilson}  &  \multicolumn{2}{c}{Kodaikanal} \\
\hline                        % inserts single horizontal line
                      &  $r_{c}$  &  $P$ &  $r_{c}$  &  $P$  \\
\hline
$\langle\alpha\rangle$                               & $-0.88$  & 0.05 & $-0.32$ & 0.48\\
$\langle\alpha_{\omega}\rangle$                      & $-0.77$ & 0.13 & $-0.57$ & 0.18 \\
\hline
$\langle\alpha\rangle/\langle\lambda\rangle$           & $-0.46$  & 0.30 & $-0.37$ & 0.41 \\
$\langle\alpha_{\omega}\rangle/\langle\lambda\rangle$  & $-0.67$ & 0.10 & $-0.61$ & 0.15 \\
\hline
\end{tabular}
\end{minipage}
\parbox[b]{9cm}{Correlation coefficients are represented by $r_{c}$ and the probability that the correlation is due to chance by $P$ for both the MW and KK data sets.}
\end{table}

In general the correlations of the strength and amplitude with the 4 tilt angle based quantities are low and inconsistent between the two data sets. Only correlations between $(\langle\alpha_{w}\rangle/\langle\lambda\rangle)_{i}$ and $L_{i+1}$ appear to be statistically significant for both data sets. Table 3 lists the correlations found between the tilt angle based parameters and the length of the next cycle. 
For MW data, the probabilities that the correlations are due to chance are below or around $10\%$ for all the averages considered except for $\langle\alpha\rangle/\langle\lambda\rangle$. The strongest correlation found is with the mean tilt angle ($\langle \alpha \rangle$) of value $r_{c}=-0.88$. In contrast, for KK data only $\langle\alpha_{\omega}\rangle/\langle\lambda\rangle$ presents a correlation ($r_{c}=-0.61$) with a chance probability below $15\%$.
This suggests that if the tilt angles are large, then the next cycle will be short. The fact that the correlation of tilt with the length of the next cycle is significant, but is poor with the strength of the next cycle is consistent with the finding that the length and strength of a cycle are poorly correlated ($r_{c}=-0.37$, \citealt{charbonneau00} and $r_{c}=-0.35$, \citealt{solanki02})

Now, tilt angles influence the amount of magnetic flux reaching the poles \citep{baumann04} and the polar magnetic flux during activity minimum has been found to be one of the proxies that best predicts the strength of the next cycle \citep{makarov89, dikpati08-b}. However, the tilt angle is not the only parameter influencing the polar flux, which is in line with the poor correlation found between tilt angles and the strength of the following cycle (values range from 0.40 to 0.54 for MW and $-0.58$ to 0.19 for KK and are not even consistent in sign between the two data sets). Obviously, the total amount of magnetic flux emerging over a cycle, $\phi_{tot}$, is another central parameter influencing the polar flux \citep{baumann04}. Hence a more physically motivated quantity to consider is $\phi_{tot}\langle\alpha\rangle/\langle\lambda\rangle$ or $\phi_{tot}\langle\alpha_{\omega}\rangle/\langle\lambda\rangle$. Since no regular and consistent magnetic information is available prior to cycle 20 we use sunspot areas as proxies of $\phi_{tot}$. I.e. instead of $\phi_{tot}\langle\alpha\rangle/\langle\lambda\rangle$ we determine $S\langle\alpha\rangle/\langle\lambda\rangle$. Sunspot areas are proportional to the amount of magnetic flux emerging through the spots since the field strength averaged over a sunspot is similar \citep{solanki03}.

Figure 5 shows $(S\langle\alpha_{w}\rangle/\langle\lambda\rangle)_{i}$ versus $S_{i+1}$. Again the errors are treated by means of error propagation and assuming Gaussian statistics and are thus overestimated. In this case we were not able to propagate the errors precisely since we have no information on the errors of the individual measured sunspot areas that would affect the calculation of $S_{i}$ or $S_{i+1}$. The error bars in Fig. 5 are calculated assuming that $S_{i}$ is known accurately. Using $S$ instead of $\phi_{tot}$ also means that we implicitely assume that the ratio of magnetic flux in faculae and network to that in sunspots is the same for each cycle. Consequently, the plotted 1$\sigma$ error bars represent lower limits to the true uncertainties. Both data sets show a moderate positive correlation between $(S\langle\alpha_{w}\rangle/\langle\lambda\rangle)_{i}$ and $S_{i+1}$ (see upper row of Table 4 for $r_{c}$ and $P$ values). It is interesting to point out that both data sets, although independent, display almost identical fits: $y_{i}=0.20 S_{i+1}+495066.6$ and $y_{i}=0.20 S_{i+1}+535359.8$ for MW and KK data sets respectiely, where $y=S\langle\alpha_{w}\rangle/\langle\lambda\rangle$.

%figura de alfa/lat * strength vs. strength +1
   \begin{figure}[h]
   \flushleft
   \includegraphics[scale=0.5]{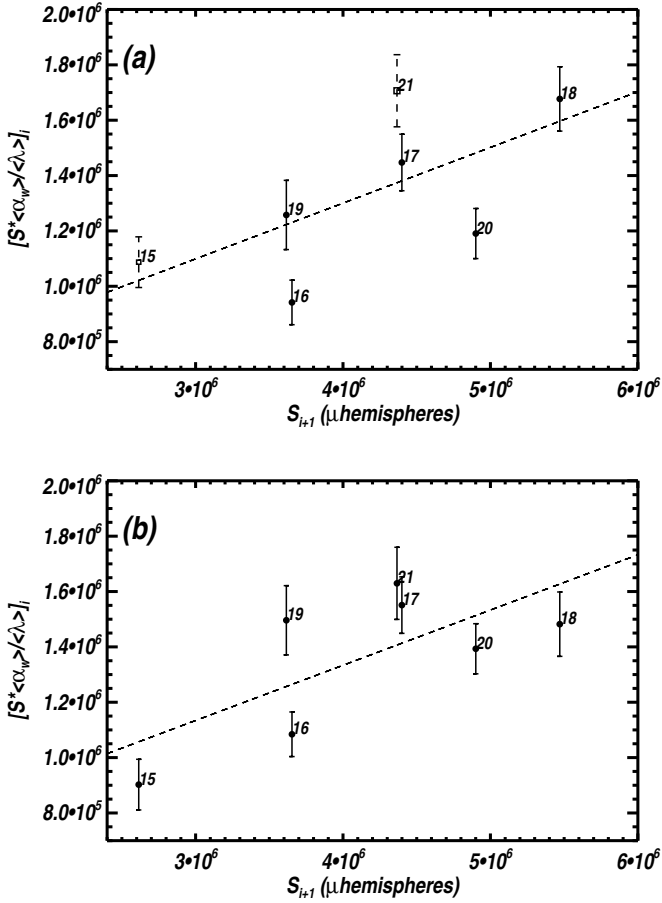}
      \caption{Strength of cycle multiplied by normalized mean area-weighted
               tilt angle vs. the strength of the next cycle for (a) MW data ($r_{c}=0.65$) and (b) KK data ($r_{c}=0.70$). The dashed lines are linear fits to the data points and the error bars represent $1\sigma$ error (assuming Gaussian statistics). For the MW data set cycles 15 and 21 are shown as squares and dashed lines for the error bars to indicate their incompleteness.
              }
       %  \label{FigVibStab}
   \end{figure}

Finally, we carry out a variant of the above analysis that is guided by dynamo models that include the influence of the meridional circulation at the solar surface \citep{wang91,choudhuri95}. According to such models the amount of flux reaching the poles depends (for a fixed differential rotation, meridional circulation and diffusion rate) on the tilt angles and the latitude distribution of the active regions. If a region is at relatively high latitude, then in general both polarities are dragged to the pole by meridional circulation, leading to a negligible change in the magnetic flux there. For active regions close to the equator the magnetic flux of the leading portion can reach and cancel the opposite polarity of the leading portion of an active region in the other hemisphere. This leads to an imbalance in the sense that mainly flux from the following polarity reaches the pole. Hence regions at low latitudes contribute disproportionately to the reversal and accumulation of magnetic field in the poles. This is thought to affect the strength of the next cycle since the polar fields are the input for the next cycle. We take into account this latitude dependence by multiplying an exponential function of the latitude to the area-weighted tilt angles. The monthly means of area and latitude weighted tilt angles are computed as follows:
\begin{displaymath}
  \overline{\alpha_{a,\lambda}}=\frac{\sum A_{j}\alpha_{j}e^{-|\frac{\lambda_{j}}{\lambda_{0}}|}} {\sum A_{j}} ,
\end{displaymath}
where $A_{j}$ is the area of the sunspot group $j$ from MW and KK data sets, $\alpha_{j}$ the tilt angle of the same group, $\lambda_{j}$ its latitude and $\lambda_{0}$ is a constant that determines how rapidly the exponential function drops with latitude. The value of $\lambda_{0}$ depends on the latitudinal velocity profile of the meridional flow, which (for reasons of symmetry) is zero at the equator. Small values of $\lambda_{0}$ correspond to a meridional flow whose horizontal component increases rapidly with $\lambda$. In the absence of clear observational constraints we have set $\lambda_{0}$ to $10^\circ$, $20^\circ$ and $30^\circ$ and have repeated the analysis for each of these values.

In Fig. 6 we plot $\overline{S}\overline{\alpha_{a,\lambda}}$ (solid curve) for the whole time series smoothed over 24 months, with $\overline{S}$ being the monthly means of sunspot area from \citet{balmaceda09}. The dashed curve is the 12 month smoothed $\overline{S}$ and the solid curve has been shifted by +11 years to better compare $\overline{S}\overline{\alpha_{a,\lambda}}$ with $\overline{S}$ of the following cycle. Since the whole curve has been shifted by a constant value and each cycle has a different duration, the lengths of the cycles of the solid and dashed curves do not match. However, it is seen that the consecutive rise in strength from cycles 16 to 19 and the drop from cycle 19 to 20 are reproduced. The correlation coefficients between the peaks of both curves reach a maximum value of 0.79 (P= 3\%) for MW and 0.78 (P=4\%) for KK when taking $\lambda_{0}=10^\circ$ (see lower row of Table 4). All of the 7 cycles are considered in both data sets since the maxima of cycles 15 and 21 are included in the MW record. The correlation coefficient values range from 0.74 to 0.79 for the MW data set and 0.78 to 0.79 for the KK data set when smoothing over 24, 36 and 48 months. Since the first 4 years of cycle 15 are not complete in the MW data set, we chose a 24 month smoothing as optimal to reduce the noise while not losing the maximum of cycle 15.

%figura de alfa/lat * strength vs. time
   \begin{figure}[h]
   \flushleft
   \includegraphics[scale=0.25]{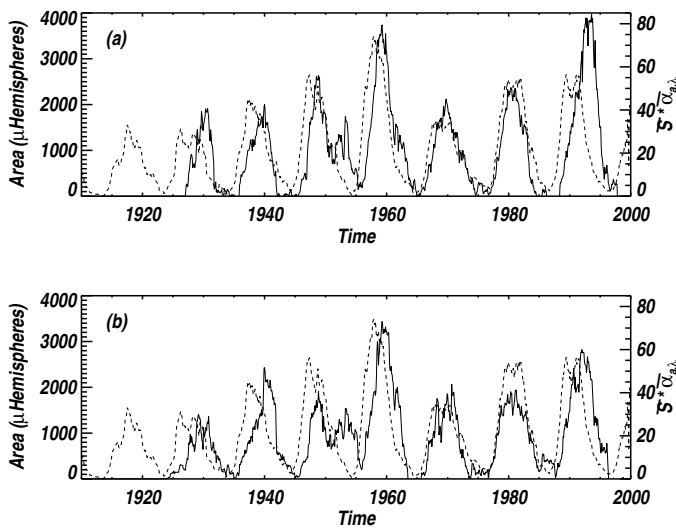}
      \caption{Comparison of actual and predicted sunspot area. The dashed curve shows monthly sunspot areas smoothed over 12 months from \citet{balmaceda09}. The solid curve is the prediction based on the tilt angles and sunspot areas of the previous cycle for (a) MW and (b) KK records, both smoothed over 24 months.}
       %  \label{FigVibStab}
   \end{figure}

 \begin{table}[h]
   \caption{Correlation coefficients between expressions containing $\alpha$ of cycle $i$ and the strength or maximum amplitude of cycle $i+1$.}
   \label{table:4}      % is used to refer this table in the text
   \centering                          % used for centering table
   \begin{tabular}{c c c| c c}  % centered columns (7 columns)
   \hline\hline                 % inserts double horizontal lines
              & \multicolumn{2}{c|}{Mount Wilson} & \multicolumn{2}{c}{Kodaikanal} \\
   \hline                        % inserts single horizontal line
                                             &  $r_{c}$  &  $P$   &  $r_{c}$  &  $P$   \\
   \hline
$(\frac{<\alpha_{w}>}{<\lambda>}S)_{i}$ vs. $S_{i+1}$      & $0.65$ & 0.11   & $0.70$   & 0.08  \\
$max(\overline{S} \overline{\alpha_{a,\lambda}})+11yrs$ vs. $max(\overline{S})$     & $0.79$ & 0.03   & $0.78$   & 0.04  \\

   \hline                                   %inserts single line
   \end{tabular}
\parbox[b]{9cm}{Correlation coefficients are represented by $r_{c}$ and the probability that the correlation is due to chance by $P$ for both the MW and KK data sets. The two rows correspond to different expressions explained in the main text.}
   \end{table}

%______________________________________________________________

\section{Discussion}

Our understanding of the solar dynamo remains incomplete despite the large
amount of effort invested into it \citep{charbonneau05}.
One hindrance to a better understanding is the limited number of observational constraints,
some of which are reviewed by \cite{gilman86, gilman02} and \cite{Rempel08}.

In this paper we have analyzed the time-dependence of the tilt angle, which has added two more
observational constraints that dynamo models must satisfy. The first is that there is an inverse 
correlation between the strength of a cycle and 
the tilt angle ($\alpha$) of sunspot groups observed during that cycle. This correlation was found to 
increase when the latitudinal dependence of Joy's law was taken into account 
(that is when $\alpha/\lambda$ was considered instead of $\alpha$). 

The results in the previous section are based on observations.
While this paper concentrates on the observational signature, it is worthwhile to 
digress and speculate on the possible causes of this inverse correlation.
In doing so we emphasise that our discussion is
only speculative whereas the results of the previous section are based firmly in the observational
data sets.
One possibility is that the field strength of magnetic flux tubes in the overshoot region
below the convection zone is larger during strong cycles.
Since stronger flux tubes are less affected by the Coriolis force this would
explain the observed correlation. In Babcock-Leighton type dynamos
the toroidal flux tubes at the base of the convection zone are believed to be the 
result of the differential rotation winding up the poloidal magnetic field. The
magnetic energy density of the loops formed in this way is likely to be limited to equipartition values with the
kinetic energy density of the differential rotation. This gives a magnetic field
strength of $\approx 10^{4}$ G. Such a loop can then loose mass via an instability 
which drains mass from the slightly sub-adiabatic region where the flux tube is located
into the covection zone, which increases the field strength to
$\approx 10^{5}$ G over a timespan of approximately 6 months \citep{rempel01}. 
As the flux tube becomes stronger, it becomes subject to the Parker 
instability which causes it to erupt to the surface. The Parker instability depends on both the field strength and
on the sub-adiabaticity of the layer where the flux tubes are stored: magnetic fields of tubes
that are stored slightly deeper can become stronger prior to the onset of the 
Parker instability.

To explain the observations requires either that in strong cycles the tubes are 
produced or stored slightly deeper \citep{caligari95}, or the region where 
the flux tubes are stored becomes slightly more stable, or that the 
intensification process acts more quickly so that higher field strengths can
be reached before the tubes erupt.  A combination of these processes may
also be at work.

There are a number of nonlinear, competing, factors which are likely to
be relevant. For example the increased flux of a strong cycle will be
more resistent to downward pumping, and will perhaps decrease the depth of the
convective overshooting at the base of the convection zone. This might decrease the 
depth at which the flux is located, but will also affect the thermodynamic
properties of the layer. Another effect, which acts in the correct direction, 
is the magnetic tension associated with the enhanced poloidal flux 
of strong cycles. This will tend to pull the field lines deeper into the 
overshoot region, 
however the effect is likely to be weak, possibly depending on how the 
poloidal field is structured.

In relation to changing the subadiabaticity, \citet{rempel01} have argued
that the energy to intensify the toroidal flux tubes to $10^{5}$G comes 
from moving material along the entropy gradient near where the tubes are 
stored. The amount of energy involved has been estimated \citep[e.g.][]{steiner04} to 
be approximately $10^{40}$ ergs. On the observational side, 
\cite{basu08} reported a $10^{-2}\%$ 
change in the wave speed squared near the base of the convection zone.
The observed change is small and anticorrelated with activity. 
Its effect in the current context is to change the subadiabaticity -- 
enhancing the stability in the region where the flux tubes are stored. 
How strong the effect would be, and how it balances with other effects, 
needs to be evaluated.

Another possibility is that the observed tilt angles have been influenced
by the near-surface flows.
These inflows consist of a time-dependent component of the solar differential
rotation \citep{howard80} called zonal flows, and a time-dependent component in the meridional plane
which has been observed by tracking magnetic features \citep{komm93} and with helioseismology 
\citep{basu03}. Some models suggest that both components of the inflow 
are driven by the excess cooling associated with plage \citep{spruit03, rempel06}.
If so, the strength of the inflow will be determined by the amount of plage, which is directly 
related to the strength of the current cycle.

Both components of the inflows will tend to decrease the tilt angle over time. 
This effect acts on the flux  tube both as it rises through 
the inflow and as it evolves on the surface after emergence. 
The sign of this effect is correct, enhanced inflows during strong cycles will reduce the observed 
tilt angles. To estimate the magnitude of the effect we now concentrate on the meridional
component of the inflow (the effect of the time variations of the zonal flows turns 
out to have a similar magnitude).

The time dependent meridional flow includes an inflow towards the
active region belts \citep{Zhao04} which had an amplitude of $\pm 5$m/s for cycle 23.
To estimate the expected magnitude of the effect on the tilt angle, 
we assume that sunspots are, on average,
subject to this flow for $\approx 5$ days before they are observed. These 5 days
include the rise time through the flow, the sunspot formation time as well as a delay caused by the
fact that not all sunspots appear on the side of the Sun facing the Earth. The maximum relative velocity with which the
two polarities could be driven towards one another by such a flow is 10m/s. This maximum is unlikely to 
be reached, so for the purposes of obtaining
a preliminary estimate we assume that they actually move towards each other with half this 
speed, that is 5m/s. 
Over the course of the 5 days this gives a 2.16Mm decrease in the latitudinal separation of their leading 
and trailing fluxes. For an active region 
with a longitudinal separation between the leading and trailing fluxes of 100Mm, this is a decrease in tilt angle of around  
1.23 degrees. If we allow for the fact that cycle 19 was stronger than cycle 23, so that its inflows would have been stronger, then the magnitude of the effect is approximately consistent with the observations.

The above are not the only possibilities for the observed negative correlation, and have been presented 
only to give an indication of some of the different types of possibilities. It is possible that
some of these explanations will be able to be excluded on either observational or theoretical grounds.
More modelling and observations will be required to pinpoint the main mechanism.

The second result is that there is a reasonably strong correlation between the product 
of the strength of a cycle and its average tilt angle and the strength of the next cycle, $r_{c}=0.65$ and $r_{c}=0.70$
for the two data sets, respectively. This product was considered because it corresponds to the poloidal source term in dynamos based on 
the Babcock-Leighton idea. The correlations were found to improve when the quantities were made to more closely match 
the poloidal source term of the models. Specifically, we found an improvement when we included a latitudinal dependence 
designed to model the effectiveness of flux emergence in producing  global poloidal fields (which depends upon some of 
the flux crossing the equator so that regions emerging close to the equator are more effective). 
This observational constraint supports the flux-transport dynamo model. It shows that 
the strength of a cycle is correlated at the 79\% or 78\% level (depending on the data set) with the poloidal flux of the previous cycle. 
Importantly, the drop in strength from cycle 19 to 20 is well reproduced. This tilt angle then appears to account for 
a substantial part of the variation from cycle to cycle of the activity level. 

This is good news in two regards. 
Firstly, it suggests that, by measuring the tilt angle and amplitude of a cycle, we will be able to make early predictions 
of the strength of the proceeding cycle. The predictive accuracy is not higher than, for instance methods based on precursors \citep[see][]{hathaway99}, but can be made much earlier. This can be seen in Figure 6
where the ``predictions'' have been shifted by eleven years (note that a two year smoothing has been performed which reduces 
the predictive horizon to 10 years). A possible improvement of predictive skill at later times might be possible
by combining different schemes, but this will depend on how independent they are.
Secondly, it suggests that a major part of the nonlinear cycle modulation is associated with 
the tilt angle. Several possible non-linearities were discussed above, such as 
the near-surface inflows, the depth at which the tubes are stored and the properties
of the plasma near the base of the convection zone. More work is required to distinguish
between these and other possibilities.

%______________________________________________________________

\section{Conclusions}

We have analyzed the sunspot data from Mount Wilson (MW) and Kodaikanal (KK) observatories in order to study Joy's law, the variation of sunspot group tilt angles from cycle to cycle and the relationship of this variation with 3 solar cycle parameters: strength, amplitude and length. The correlations found from the analysis are listed in Tables 1, 2, 3 and 4. From the analysis we highlight the following:\\

(1) A linear fit to Joy's law gives $\alpha=(0.26\pm0.05)\lambda$ for the MW and $\alpha=(0.28\pm0.06)\lambda$ for the KK data sets. Here $\alpha$ is the tilt angle and $\lambda$ the latitude, both expressed in degrees.

(2) The mean tilt angle changes from cycle to cycle (Fig. 2a and Table 1). The range of values exceeds the uncertainties in the cycle-averaged tilt angles.

(3) A negative correlation between the normalized tilt angle, or $\langle\alpha\rangle/\langle\lambda\rangle$, and the strength of the same cycle is found ($r_{c}=-0.95$ and $r_{c}=-0.93$ for MW and KK data sets, respectively).

(4) We also find a negative correlation between the latitude normalized area weighted tilt angle ($\langle\alpha_{w}\rangle/\langle\lambda\rangle$) and the length of the next cycle ($r_{c}=-0.67$ and $r_{c}=-0.61$ for MW and KK data sets, respectively).

(5) Finally, we discovered a positive correlation between the strength of one cycle multiplied by its mean area- and latitude- weighted tilt angle, $(S\langle\alpha_{w}\rangle/\langle\lambda\rangle)_{i}$, and the strength of the  \textit{next} cycle, $S_{i+1}$, ($r_{c}=0.65$ and $r_{c}=0.70$ for MW and KK data sets, respectively). Higher correlation coefficients are obtained between a tilt-angle based expression obtained through guidance from Babcock-Leighton type dynamo models and the amplitude of the next cycle ($r_{c}=0.79$ and $r_{c}=0.78$ for MW and KK, respectively).\\

These results show the importance of the tilt angle of sunspot groups for both the prediction of solar activity and the understanding of the physics behind the solar dynamo.

%______________________________________________________________

\begin{acknowledgements}
      This work was supported by the German
      \emph{Deut\-sche For\-schungs\-ge\-mein\-schaft, DFG\/} project
      number SO-711/1-3, and by the WCU grant No. R31-10016 funded by the Korean Ministry of Education, Science and Technology.
\end{acknowledgements}

 \bibliographystyle{aa.bst}
 \bibliography{marybib.bib}

\begin{appendix}
\section{Determining the cycle-to-cycle variaitons in the presence of a large intrinsic scatter.}

As has been metioned, the tilt angles of individual active regions are largely random, with Joy's
law being a relatively small bias.  The purpose of this appendix is to discuss the calculation of 
cycle-to-cycle changes in Joy's law from the data. For this purpose we will assume that Joy's law 
applies for each cycle and that the scatter in the data is random and unbiased. More explicitly we assume
\begin{enumerate}
\item{the tilt angle, $\alpha_i$,  for each spot, $i$,  obeys Joy's law 
\begin{equation}
\alpha_i= a_n \lambda_i +\epsilon_i
\end{equation}
where $a_n$ is the (possibly cycle-to-cycle dependent) constant of proportionality for cycle $n$
and $\epsilon_i$ represents the random deviation from Joy's law of individual sunspot 
groups.}
\item{The $\epsilon_i$ are independent realizations of a random process with a mean of zero.}
\end{enumerate}

We calculate our estimate, $b_n$, of $a_n$ for each cycle according to 
\begin{equation}
b_n=\frac{\sum_{i} \alpha_i}{\sum_{i} \lambda_i}
\end{equation}
where the sum is again over spots in cycle $n$. The error of the approximation is 
\begin{equation}
b_n-a_n=e_n=\frac{\sum_{i} \epsilon_i}{\sum_{i} \lambda_i}.
\end{equation} 

We also consider $\bar{a}$, the value of $a$ based on the whole data set, ignoring cycle-to-cycle changes in $a$. 
The equivalent estimate, $\bar{b}$ of $\bar{a}$ has the summation extended to all cycles.  
Note that we could have also considered the cycle-to-cycle deviation of Joy's
law, $d_n$, from the estimate obtained over all cycles $\bar{b}$. This however
has exactly the same error as does $a_n$ as can easily be seen:
\begin{eqnarray}
d_n&=&\frac{\sum_i \alpha_i -\bar{b} \lambda_i}{\sum_i \lambda_i}\\
d_n&=&\frac{\sum_i a \lambda_i +\epsilon_i -\bar{b} \lambda_i}{\sum_i \lambda_i} \\
%d_n&=& a+\frac{\sum_i \epsilon_i}{\sum_i \lambda_i}-\bar{b}\\
d_n&=&a-\bar{b}+\frac{\sum_i \epsilon_i}{\sum_i \lambda_i}\\
d_n&=&a_n-\bar{b}+e_n
\end{eqnarray}
This result indicates that calculating the cycle-to-cycle deviations from a reference Joy's
law is identical to calculating Joy's law for each cycle and then subtracting a fixed
constant.
A similar result can be shown if $\epsilon_i$ is assumed to be dependent on the area,
as was considered in the main text.
\end{appendix}

\end{document}